\documentclass[journal]{IEEEtran}
\usepackage{cite}
\usepackage{setspace}

\ifCLASSINFOpdf
  \usepackage[pdftex]{graphicx}
  \graphicspath{{../pdf/}{../jpeg/}}
  \DeclareGraphicsExtensions{.pdf,.jpeg,.png,.bmp}
\else
\fi
\usepackage{amsmath}
\usepackage{array}
\usepackage{multirow}
\usepackage{pbox}

\ifCLASSOPTIONcompsoc
  \usepackage[caption=false,font=normalsize,labelfont=sf,textfont=sf]{subfig}
\else
  \usepackage[caption=false,font=footnotesize]{subfig}
\fi
\begin{document}
\title{A Robust Blind Watermarking \\ Using Convolutional Neural Network}

\author{Seung-Min Mun, Seung-Hun Nam, Han-Ul Jang, Dongkyu Kim, and Heung-Kyu Lee%,~\IEEEmembership{~Senior member,~IEEE}% <-this % stops a space
%\thanks{M. Shell was with the Department
%of Electrical and Computer Engineering, Georgia Institute of Technology, Atlanta,
}


\markboth{}%IEEE SIGNAL PROCESSING LETTERS,~Vol.~14, No.~8, August~2015
{Shell \MakeLowercase{\textit{et al.}}: Bare Demo of IEEEtran.cls for IEEE Journals}



\maketitle

\begin{abstract}
This paper introduces a blind watermarking based on a convolutional neural network (CNN). We propose an iterative learning framework to secure robustness of watermarking. One loop of learning process consists of the following three stages: Watermark embedding, attack simulation, and weight update. We have learned a network that can detect a 1-bit message from a image sub-block. Experimental results show that this learned network is an extension of the frequency domain that is widely used in existing watermarking scheme. The proposed scheme achieved robustness against geometric and signal processing attacks with a learning time of one day.
\end{abstract}

\begin{IEEEkeywords}
digital watermarking, color image watermarking, blind watermarking, convolutional neural network(CNN).
\end{IEEEkeywords}

% \fi
\IEEEpeerreviewmaketitle

\section{Introduction}
\IEEEPARstart{W}{atermarking} is used to identify and protect ownership of copyrighted media content, by embedding invisible data into the original content. The most difficult requirements for watermarking schemes is robustness; the detector should be able to extract the watermark properly even if the content has little distortion. The second requirement is invisibility; the visual quality should not be damaged too much by the mark.

%In this paper, we optimize robustness while considering invisibility.

Over the last decade, most watermarking techniques have acquired robustness using the frequency domain. Generally, the first step is to use a transform such as DCT or DWT  to create a set of values corresponding to the original pixel data. Then some of the values in the transformed domain are modified and become the pixel data by inverse transform \cite{kang2003dwt,lin2000robust}. In recent years, quaternion discrete Fourier transform (QDFT) performs best for blind watermarking techniques for color images as shown in \cite{wang2013robust,chen2014full,ouyang2015color}. These techniques are robust against signal processing attacks by applying QDFT for small image blocks. For geometric attacks, the robustness was secured with the help of a template. However, since their templates are limited to compensating for RST attacks, they are vulnerable to general affine transformation.

\begin{figure}[!t]
\centering
\includegraphics[width=3.3in]{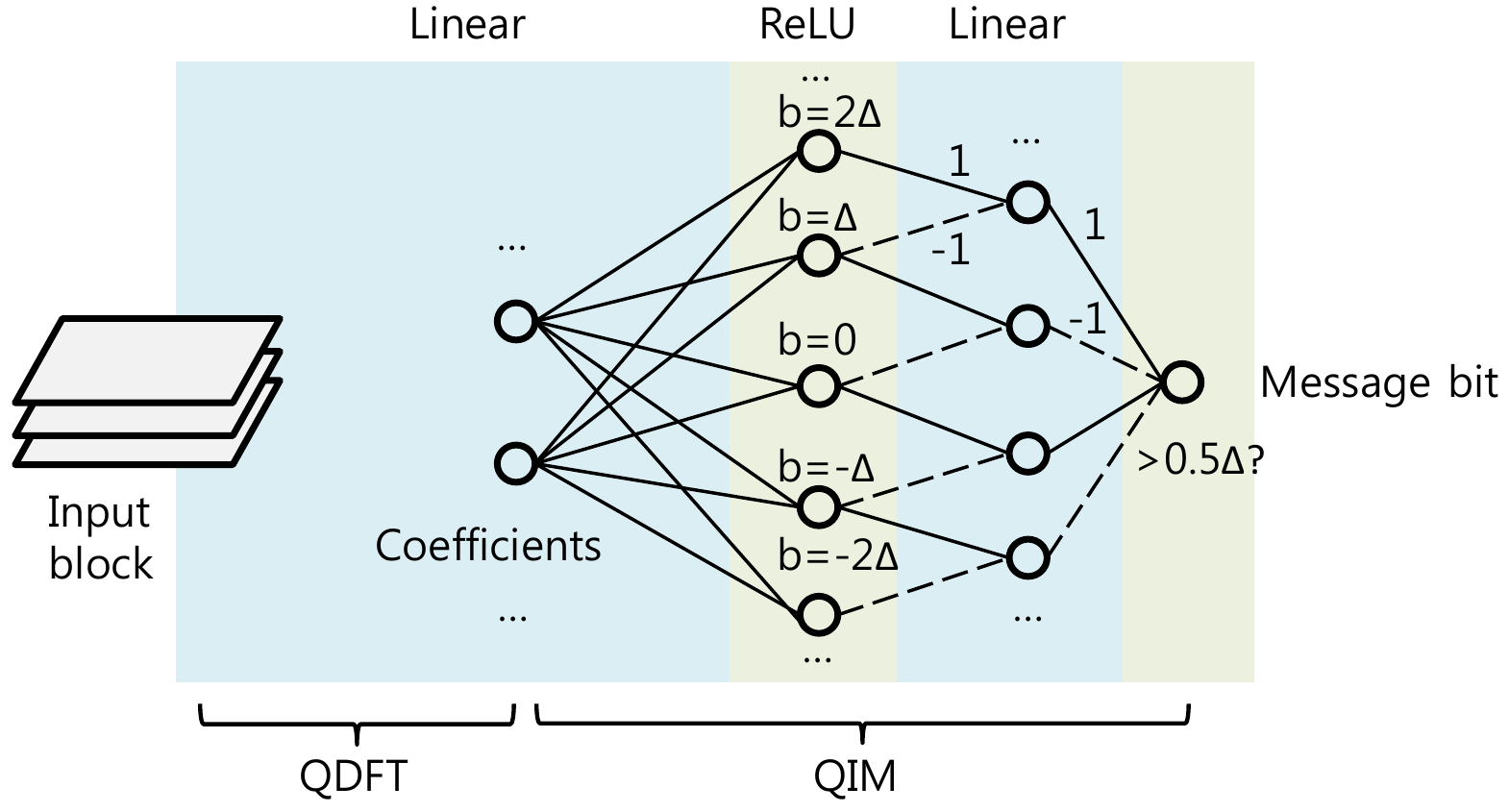}
\caption{A network representation of an existing watermarking domain. The same logic holds for a linear transform such as DFT or DCT as well as QDFT. Where $\bigtriangleup$ is the quantization step of QIM.}
\label{layers}
\end{figure}

In this paper, we optimize robustness while considering invisibility in depth. Our goal is to learn convolutional neural network (CNN) beyond QDFT and replace it. In fact, watermark detection using QDFT is representable by a very shallow network. According to \cite{chen2014full}, each component of QDFT can be obtained as a linear combination of DFT coefficients on each color channel. Most of recent techniques use quantization index modulation (QIM)\cite{wang2013robust,chen2014full,ouyang2015color}, which can be represented by one ReLU and two fully connected layer back and forth. Here, adjacent linear layers can be replaced by one linear layer. Taken together, QDFT and QIM is like combining three layers in a linear, ReLU, and linear order as shown in Figure \ref{layers}. Therefore, CNN with a sufficiently large number of layers and filters can be seen as a generalization of QDFT and QIM. The proposed CNN model has 12 ReLU layers alternating with linear layers. In addition, the proposed model parameters are optimized for learning while QDFT is fixed. Unlike QDFT, the proposed scheme has domain specialized to watermarking.
%  It is possible that the proposed CNN model imitates or surpasses QDFT.

\begin{figure}[!t]
\centering
\includegraphics[width=3in]{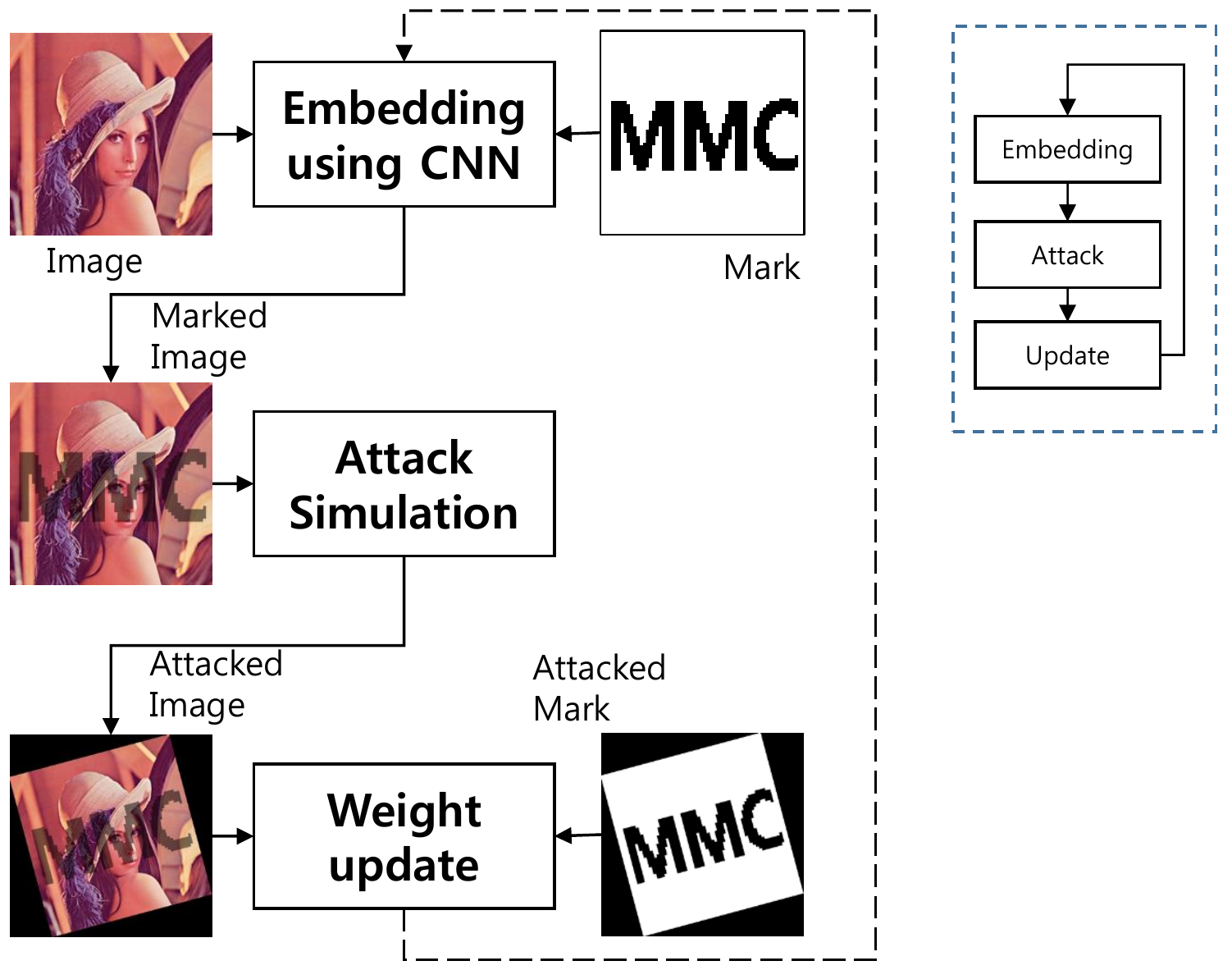}
\caption{Framework of learning CNN for watermarking. The embedded mark is shown as visible for better understanding. At the top right corner is a simplified illustration}
\label{flowchart}
\end{figure}

As shown in Figure \ref{flowchart}, we propose a light-weight and simple framework using a CNN. And this iterative process obtains a robust domain based on the given attack such as JPEG, resizing and noise addition. As a result, when new attacks are given, our framework can adaptively re-acquire robustness.

Signal processing attacks and geometric attacks were tested experimentally. The comparison with the QDFT-based watermarking paper\cite{ouyang2015color} are shown. Experimental results show that the learning-based domain can surpass the existing frequency domain used for watermarking.

\section{Proposed CNN based watermarking}
\subsection{Main idea of learning framework}
The goal of this iterative process is to construct a CNN model that takes an $R\times C$ size block as input and determines a message bit. This CNN is then used in a similar way to QDFT in traditional frequency domain watermarking techniques \cite{wang2013robust}. Training comprises three stages starting from a CNN with random weights. This three stages are repeated until the watermark that CNN embedded is correctly identified after the attack. In this subsection we provides details for each stage.

%\begin{figure}[]
%  %\caption{A picture of a gull.}
%  %\centering
%\includegraphics[width=3cm]{flowchart.pdf}
%\label{vis}
%\end{figure}

\subsubsection{Watermark embedding using CNN}
%백프로파게이션때 global stats을 씀
The first stage is the process of inserting a watermark into the image using the CNN of the previous loop. This stage accepts images and watermarks of $M\times N$ size and generates watermarked images. First, images and the corresponding watermarks are divided into non-overlapping blocks of $R\times C$ size. The message bit $m$ of each block $\mathbf{B}$ is determined by looking at the watermark image of the corresponding block position. We want to incrementally change the image so that CNN recognizes block $\mathbf{B}$ as $m$. We update the image by taking the gradient decent method (SGD), which is designed to update the weights:
\begin{equation}
\mathbf{B}^{(t+1)} = \mathbf{B}^{(t)} - \alpha \nabla_\mathbf{B} L(\mathbf{W},\mathbf{B}^{(t)},m)
\label{embed}
\end{equation}
\begin{equation}
L(\mathbf{W},\mathbf{B},m)=-log\: p(m|\mathbf{W},\mathbf{B}) + \frac{1}{2}\lambda \| \mathbf{B} -  \mathbf{B}_0 \|_{2}^2
\label{loss_eq}
\end{equation}
where $\mathbf{B}^{(t)}$ is the block to embedding at iteration $t$, $\alpha$ is the embedding rate, $\mathbf{W}$ is the weights of the CNN, $L$ is the loss function and includes a cross-entropy loss term as in Equation \ref{loss}. The loss function was devised to protect the invisibility while embedding the watermark. $p(m|\mathbf{W},\mathbf{B})$ is the probability that message $m$ is inserted into $\mathbf{B}$ predicted by CNN. We introduced the $l^2$ regularization term $\| \bullet \|_2$ to ensure invisibility.

\begin{figure}[]
\centering
\includegraphics[width=3.5in]{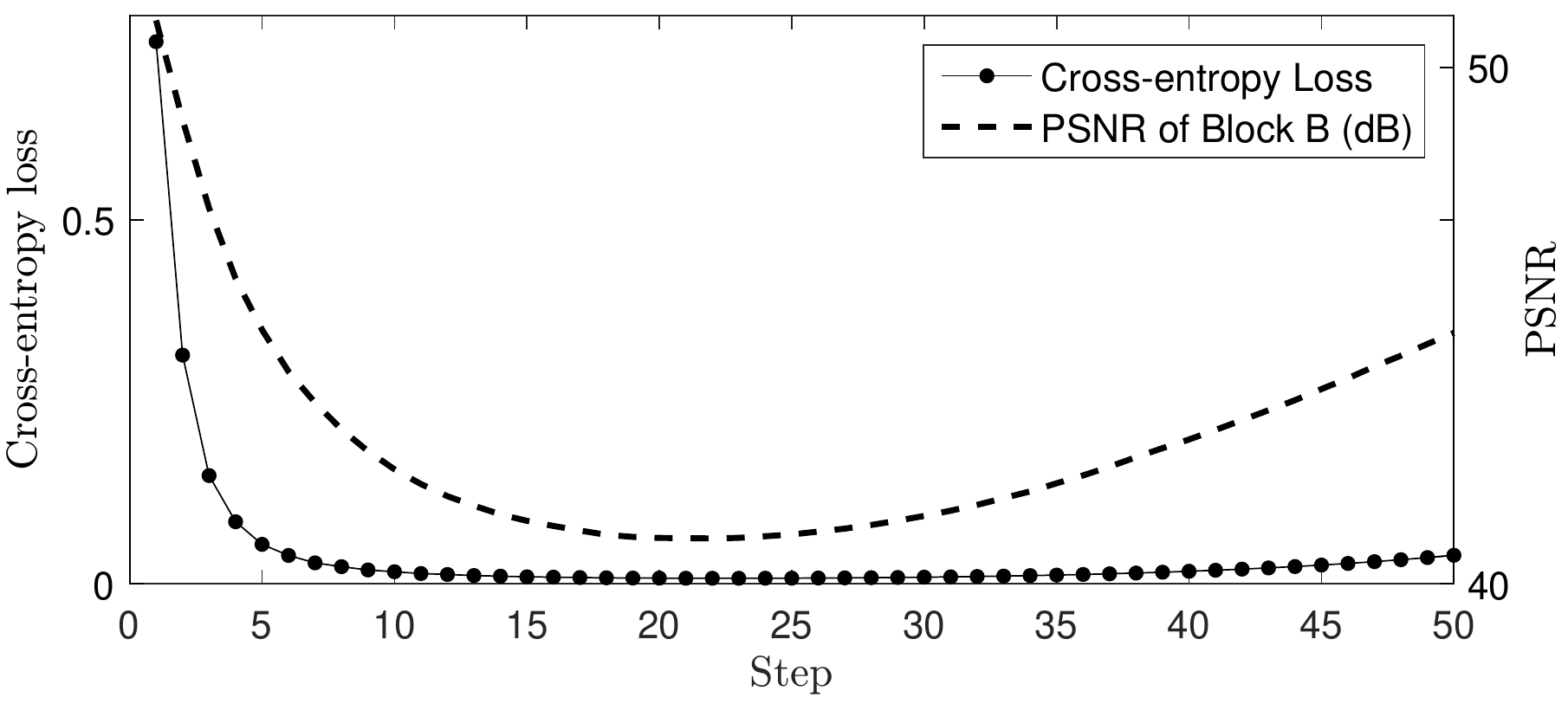}
\caption{Changes in Cross-entropy loss and invisibility according to the embedding step. The points are determined by the average of the 128 blocks. A description of the PSNR is given in Section \ref{exp_results}.}
\label{emstep}
\end{figure}

Empirically, we have confirmed that if $\alpha$ is sufficiently small, the loss gradually decreases as shown in Figure \ref{emstep} even in the case of random weights. A loss close to 0 means that the message $m$ is properly inserted into block $\mathbf{B}$. The blocks are then combined to produce a watermarked image. Note that the weight $\mathbf{W}$ does not change in this stage.

\subsubsection{Attack simulation}
Attack simulation is necessary for CNN to adaptively capture invariant features for various attacks. This stage accepts the watermarked image to produce the attacked image. The important thing is that not only the watermarked cover image but also the watermark images are attacked. This attacked watermark is used as a true label for supervised learning in the next stage. The simulated attack sets include affine transform, cropping, JPEG compression, noising, Gaussian filtering, median filtering, rotation, rescaling. For each image and watermark, a pair of attacked images is created as many times as the number of attacks.

\subsubsection{CNN weight update}
Now we update the CNN weights so we can correctly extract the watermark from the given image. That is, CNN itself has to imitate the role of the detector. This stage begins by setting the size of all the attacked image to $M\times N$. As before, divide the image into non-overlapping blocks of $R\times C$ size. Again, The message bit $m'$ of each distorted block $\mathbf{B}'$ is determined by looking at the watermark image of the corresponding block position. If the average value of the watermark image block is grater than 0.5, we set the message bit $m'$ as 1 and set as 0 otherwise. We applied the gradient descent method so that when CNN takes $\mathbf{B}'$ as an input, it correctly predicts $m'$.
\begin{equation}
\mathbf{W}^{(t+1)} = \mathbf{W}^{(t)} - \eta \nabla_\mathbf{W} L(\mathbf{W}^{(t)},\mathbf{B}',m'),
\end{equation}
where $\mathbf{W}^{(t)}$ is the weights at iteration $t$, $\eta$ is the learning rate. The loss function, $L$, is identical to that of the embedding stage.
\begin{figure}[!t]
\centering
\includegraphics[width=3.4in]{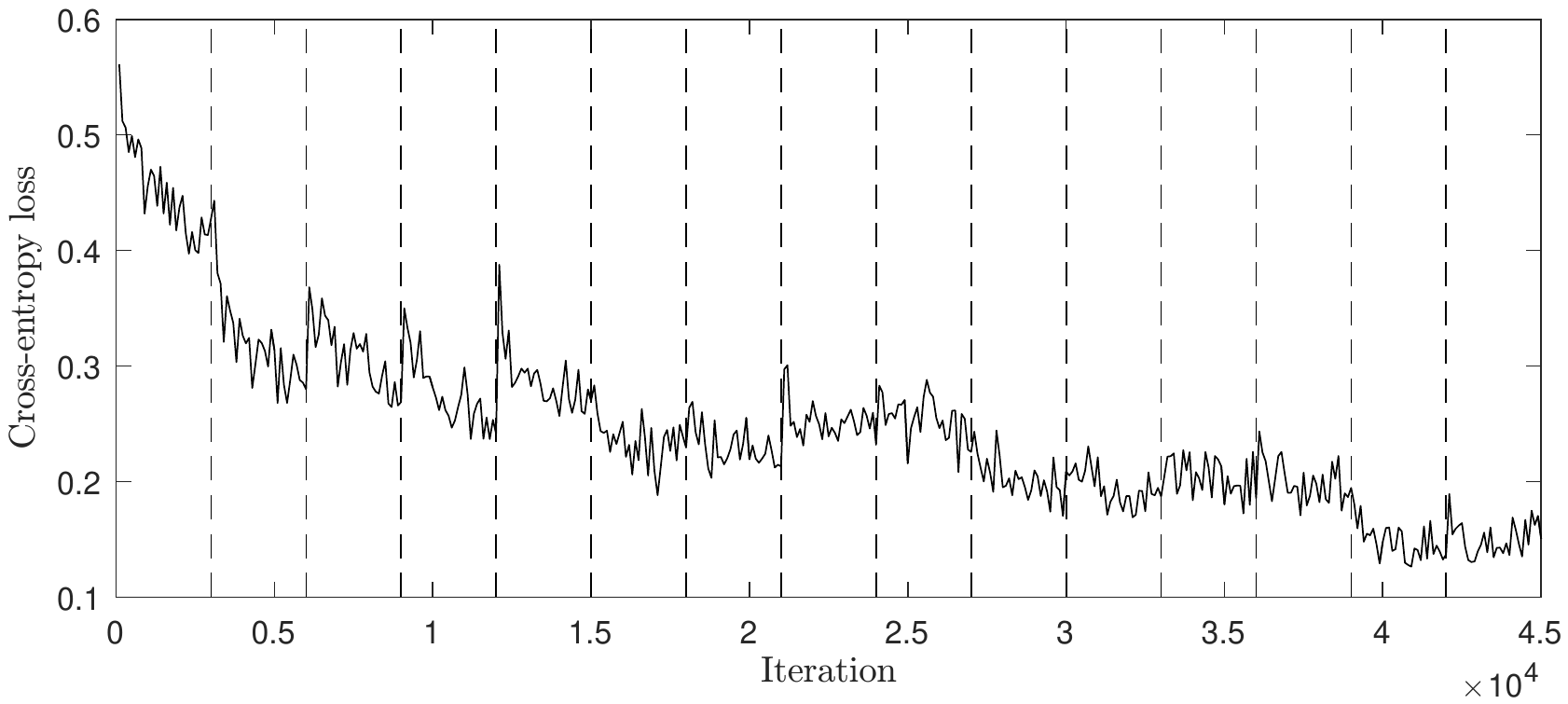}
\caption{Cross-entropy loss in the updating stage. The vertical lines divide each update stage and indicate that it has passed the embedding and attack stages.}
\label{loss}
\end{figure}
At the end of this stage, all watermarked and attacked images created so far are deleted and the process returns to the embedding stage.

%\begin{figure}[!t]
%\centering
%\includegraphics[width=3in,height = 3in]{modelblock}
%\caption{The residual block used in the proposed model, inspired by the existing residual network. Channel concatenation with the raw data has been added. The proposed model includes this block repeatedly four times.}
%\label{modelblock}
%\end{figure}
The lower the loss, the more CNN can correctly extract message bits regardless of attack. In other words, robustness against various attack is secured. We measured the loss every time the stage was repeated. Figure \ref{loss} shows empirical reduction of the loss by the proposed framework.

\subsection{Use of trained CNN}

If CNN's weights parameters converged sufficiently due to the learning framework, this subsection describes how CNN can be used for embedder and detector of applicable watermarking. The roles and functions of the embedder are described in detail in the previous subsection. The embedding process for learning and the actual application are exactly the same. However, the CNN used in this case has a weight obtained at the end of learning. The CNN as a detector extracts a 1-bit message for each $R \times C$ block as before. The CNN predicts the message from the block of the attacked image. If the probability of 0 as a result of the softmax layer is larger than 0.5, the message of the block is decided to 0. Otherwise, the message bit is decided to 1. As a result, a watermark of $(M / R) \times (N / C)$ size is extracted for one image.

\subsection{Implementation details}
%l2 norm regul === block diff decay
We propose a CNN model modified from a part of the most powerful model called Residual Network \cite{he2016deep}, which includes residual units and batch normalization. The proposed CNN contains 14 convolution layers and the kernel size of each convolution layer is 1$\times$1 or 3$\times$3. The number of filters in each convolution layer was 128. A soft-max layer is added at the end for classification. Generally, the more layers, the better the detection performance, but there is a trade-off where the feed-forwarding and the back-propagation time increase. The number of layers and filters was determined experimentally. Dropout was not used because it can be replaced by batch normalization. Pooling is aimed at ensuring spatial invariance, but it causes performance degradation to capture fine patterns. So we did not use max-pooling or average-pooling.

The image data used for training consisted of 4000 24-bit colour images from the BOSSBase dataset designed for data hiding studies\cite{bas2011break}. Without loss of generality, we set $M=N=512$ and $R=C=8$. Initially, $\alpha$ and $\eta$ were set to $0.01$ and $0.001$ respectively. We used 8 randomly sampled images per embedded stage and 15 types of attacks are used. For one image, we used a pair of watermarks, 0 and 1 being inverted from each other. Each image is divided into $4096$ blocks so $8 \times 2 \times 15 \times 4096$ blocks are used in one updating stage. The update unit, batch, consists of 128 blocks and is sampled from the attacked images.

In practice, we used adaptive moment estimation (Adam) instead of SGD because Adam is more complicated but has better loss reduction. Also, the gradient in Equation \ref{embed} is normalized with mean and standard deviation for invisibility and fast convergence. Inspired by learning rate annealing, we reduced the $\alpha$ by $0.9$ times per iteration. Iteration  $t$ for each blocks in the embedding phase is limited to $12$.

\section{Experimental results}
\label{exp_results}

The network was trained using a GPU, NVIDIA GTX 1070 for one day, and test images and test watermarks were not used for the training. Attack experiments were performed using StirMark benchmark\cite{petitcolas1998attacks} for JPEG, noising and affine transforms. For the remaining attacks, Python libraries scikit-image \cite{van2014scikit} and scipy \cite{jones2014scipy} were used.

We conducted robustness testing for signal processing attacks and geometric attacks. The comparison technique is QDFT based watermarking. One bit is inserted into the QDFT mid-frequency coefficients of four components by QIM method. Existing techniques\cite{wang2013robust, ouyang2015color} estimate the rotation angle, the moving distance, and the scaling factor at the detection step using the inserted template. We assumed that the RST correction through this estimation was performed without error to compare performance with existing QDFT, as \cite{chen2014full} did. Table \ref{table_result} shows the result of registering an attacked image first, and we implemented the comparison technique. To make the embedding capacity the same as the proposed technique, the comparison technique inserts one bit in one block as \cite{ouyang2015color} does. For robustness comparison, both techniques omitted the scrambling of the watermark.

\begin{table*}[]

\renewcommand{\arraystretch}{1.4}
\caption{Comparison of the extracted watermarks with registration}
\label{table_result}
\centering
\setlength\tabcolsep{5.0pt}
\footnotesize
\begin{tabular*}{\textwidth}{c c|c c c|c c c|c c c|c c c|c c c|}

\multicolumn{2}{c|}{\multirow{2}{*}{Attacks} }
& \multicolumn{3}{c|}{QDFT\cite{ouyang2015color}}
& \multicolumn{3}{c|}{Training A1}
& \multicolumn{3}{c|}{Training A1-A8}
& \multicolumn{3}{c|}{Training A1-A16} % A1-A13
\\
&& Baboon & Lenna & Peppers% & BOSSBase test set
& Baboon & Lenna & Peppers% & BOSSBase test set
& Baboon & Lenna & Peppers% & BOSSBase test set
& Baboon & Lenna & Peppers% & BOSSBase test set
\\[2pt]\cline{1-14}

A1&No attack (PSNR)&
35.7dB&37.2dB&37.4dB&
38.9dB&38.9dB&39.0dB&
35.1dB&38.3dB&38.8dB&
35.9dB&39.2dB&38.9dB
\\
A2&JPEG 80&
0.5008&0.7673&0.8313&
-0.0372&0.1326&0.2127&
0.5305&0.5826&0.6578&
0.6384&0.8963&0.8938
\\
A3&JPEG 90&
0.6880&0.8143&0.8850&
-0.0252&0.1886&0.2253&
0.7684&0.8571&0.9251&
0.9129&0.9676&0.9614
\\
A4&Median filtering&
0.6596&0.9460&0.9760&
0.1951&0.3221&0.3322&
0.7189&0.8461&0.8751&
0.8377&0.8987&0.8985
\\
A5&Gaussian filtering&
0.4746&0.7933&0.8947&
0.4374&0.4631&0.6218&
0.9076&0.9238&0.9454&
0.9625&0.9343&0.9590
\\
A6&Affine 1&
0.3062&0.3148&0.3229&
0.1955&0.2459&0.3128&
0.9673&0.9831&0.9910&
0.8971&0.9738&0.9451
\\
A7&Affine 2&
0.3036&0.3143&0.3201&
0.2281&0.2330&0.2482&
0.9857&0.9858&0.9933&
0.8428&0.9417&0.9236
\\
A8&Affine 3&
0.1402&0.0477&0.0185&
0.1016&0.3122&0.4088&
0.8381&0.8918&0.9005&
0.9532&0.9032&0.9090
\\
A9&Affine 4&
0.1212&0.0803&-0.0105&
0.1019&0.3083&0.3735&
0.7197&0.7952&0.8446&
0.9207&0.8343&0.8794
\\
A10&Affine 5&
0.1324&0.0899&-0.0217&
0.0239&0.2248&0.3587&
0.8093&0.8654&0.8822&
0.9450&0.9055&0.9125
\\
A11&Affine 6&
0.1494&0.0753&0.0057&
-0.0257&0.1149&0.2585&
0.7785&0.7996&0.8744&
0.9337&0.8631&0.8661
\\

A12&Noising&
0.9220&0.8880&0.9077&
0.9704&0.9863&0.9962&
0.6594&0.7167&0.9076&
0.9737&0.8784&0.9084
\\
A13&Resizing 75\%&
0.8425&0.9539&0.9780&
0.7081&0.8002&0.8360&
0.9903&0.9934&0.9970&
0.9891&0.9947&0.9961
\\
A14&Rotation 10$^\circ$&
0.6721&0.8733&0.9252&
0.5918&0.6425&0.7367&
0.9857&0.9934&0.9970&
0.9749&0.9781&0.9817
\\
A15&Rotation 90$^\circ$&
1.0000&1.0000&1.0000&
0.9967&1.0000&1.0000&
1.0000&1.0000&1.0000&
1.0000&1.0000&1.0000
\\
A16&Cropping 80\% &
0.8362&0.9910&0.9745&
0.8740&0.9868&0.9910&
0.8746&0.9921&1.0000&
0.8730&1.0000&0.9921
\\

\end{tabular*}
\end{table*}

\begin{figure}[]
\centering

\subfloat{\centering\includegraphics[width=0.12\textwidth]{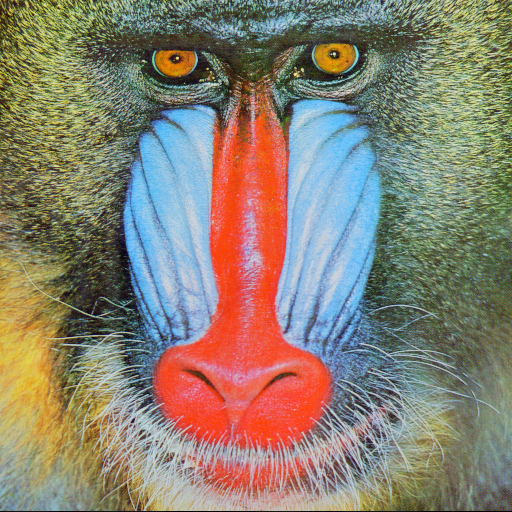}}\quad
\subfloat{\centering\includegraphics[width=0.12\textwidth]{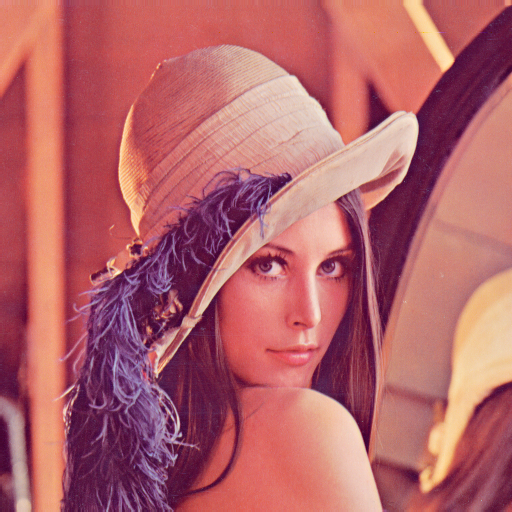}}\quad
\subfloat{\centering\includegraphics[width=0.12\textwidth]{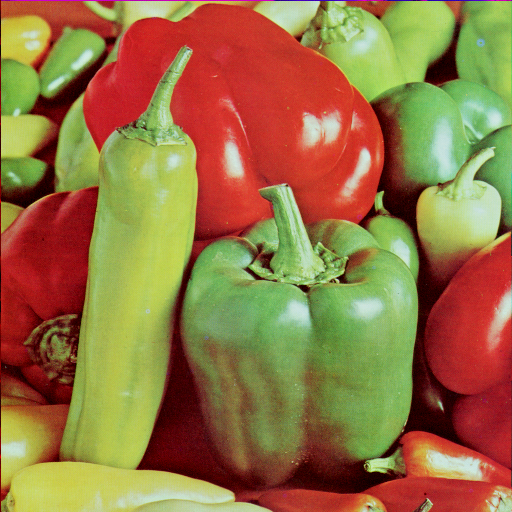}}\quad

\subfloat{\centering\includegraphics[width=0.12\textwidth]{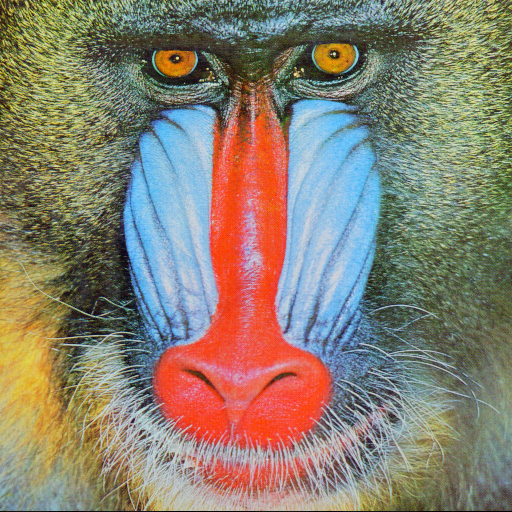}}\quad
\subfloat{\centering\includegraphics[width=0.12\textwidth]{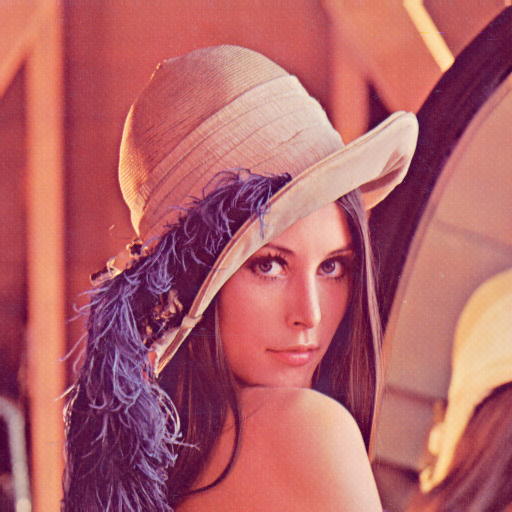}}\quad
\subfloat{\centering\includegraphics[width=0.12\textwidth]{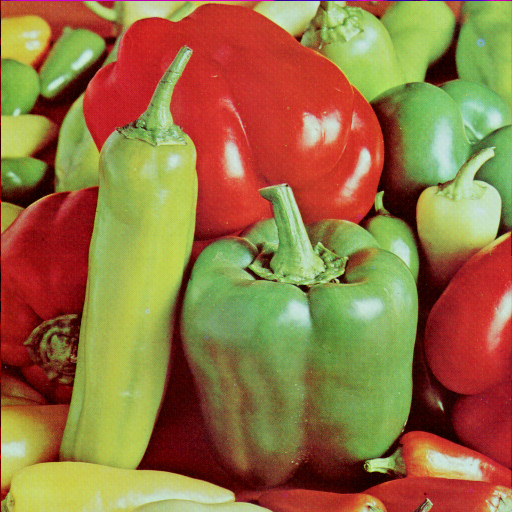}}\quad

%\caption{Attacks used for training and testing. (a) no attack, (b) median filtering, (c) JPEG 80, (d) JPEG 90, (e) noising, (f) Gaussian filtering, (g)-(l) affine transformation, (m) 10$^\circ$ rotation, (n) 90$^\circ$ rotation, (o) resizing to 75\%, (p) 80\% cropping}

\caption{Visual impact comparison before and after watermark embedding. The top row is the original test images and the bottom row is the watermarked test images.}
\label{visual}
\end{figure}

%Figure \ref{fig_attack} shows the attacks applied for testing, and these attacks were also included during training. 
Like the related works \cite{wang2013robust,chen2014full,ouyang2015color,kandi2017exploring}, standard test images Baboon, Lenna and Peppers were used as cover images. They are all 24 bit color images and are 512$\times$512 in size as shown in Figure \ref{visual}. MPEG-7 CE Shape-1\cite{latecki2002shape}, a binary image data set, was processed for use as a watermark image.

The standard deviation of the kernel used in Gaussian filtering is 1. Median filtering was performed for each channel with a 3$\times$3 kernel. We experimented with six types of affine transformations that move the pixel coordinates a bit as in Stirmark benchmark\cite{petitcolas1998attacks}:
\begin{equation}
\begin{bmatrix}
    x'\\ y'
\end{bmatrix}
=
M_{i}
\begin{bmatrix}
    x\\ y
\end{bmatrix}
\label{affine_eq}
\end{equation}
\begin{align*}
M_1&=
\begin{bmatrix}
1 & 0\\0.01 & 1
\end{bmatrix},
&M_2&=
\begin{bmatrix}
1 & 0.01\\0 & 1
\end{bmatrix},\\
M_3&=
\begin{bmatrix}
1 & 0.01\\0.01 & 1
\end{bmatrix},
&M_4&=
\begin{bmatrix}
1.01 & 0.013\\0.009 & 1.011
\end{bmatrix},\\
M_5&=
\begin{bmatrix}
1.007 & 0.01\\0.01 & 1.012
\end{bmatrix},
&M_6&=
\begin{bmatrix}
1.013 & 0.008\\0.011 & 1.008
\end{bmatrix}
\end{align*}
where $M_i$ is a matrix corresponding to transformation number $i$. $x$ and $y$ are the pixel coordinates before the transformation, and $x'$ and $y'$ are the coordinates after the transformation.

Robustness and invisibility performance were measured by normalized correlation (NC) and peak signal-to-noise ratio (PSNR), respectively.
\begin{equation}
PSNR(\mathbf{I},\mathbf{I}') =10 \log_{10}\{\frac{255^2 \times 3MN}{\| \mathbf{I} -  \mathbf{I}' \|_{2}^2}\}
\label{PSNR}
\end{equation}
\begin{equation}
NC(\mathbf{w},\mathbf{w}') =\frac{\langle\mathbf{w} , \mathbf{w}'\rangle}{\| \mathbf{w}\|_{2} \|\mathbf{w}'\|_{2}} %\middle/ () \right.
\label{NC}
\end{equation}
where $\mathbf{I}$ and $\mathbf{I}'$ denote the original cover image and the cover image after embedding, respectively. $\mathbf{w}$ and $\mathbf{w}'$ denote the bit sequences of the original watermark and the watermark detected after the attack, respectively. $\langle\cdot,\cdot\rangle$ means the inner product.
NC and PSNR are the most widely used performance indicators as shown in previous watermarking papers\cite{lin2000robust,ouyang2015color,kandi2017exploring}. 
%PSNR was displayed instead of NC in Table 1 because the watermark was detected without an error when there was no attack.

Table \ref{table_result} shows the test results according to the attack set to be trained. Two observations are derived from the third column. First, the attacks that are included in the training set show higher robustness than those that are not. Second, The robustness of the third column tends to be lower than the robustness of the rightmost column. We infer the cause of this anomaly as certain attacks will help to find robust features for other attacks when added to training. Based on two observations, it is obvious that we have to train enough of the various sets of attacks. In this paper, therefore, the rightmost column is proposed.
%The proposed scheme does not include registration and its purpose is to replace QDFT with CNN.

%Therefore, the comparative experiment was compared with the version containing the registration and the version without the registration.

%Originally, [?] Inserts each bit into four coefficients in the frequency domain of one block. To fit the insertion capacity of the comparison technique, one bit was inserted into the average of the four coefficients. As a result, the capacity of the comparison technique is reduced to $1/4$, which is equal to the capacity of the proposed technique.
\section{Discussion and related work}
The proposed scheme assume that it is used with existing template matching methods. So we can not get robust correlation results against rotation and cropping without assuming registration. The extracted watermarks in the absence of registration are as shown in table \ref{table_result_wo_reg}. The computational complexity of proposed method is higher than that of QIM after QDFT and the time required for detection is longer. Experimentally, the proposed technique took about $1.6$s to detect watermark from one image and the comparison technique took $0.5$s using a CPU i7-4770k. %Experimentally, the proposed technique took $1.6338$ seconds to detect watermark from one image and the comparison technique took $0.4926$ seconds using a CPU i7-4770k.

\begin{table}[]
\renewcommand{\arraystretch}{}
\caption{Comparison of the extracted watermarks without registration}
\label{table_result_wo_reg}
\setlength\tabcolsep{1pt}
\centering
\begin{tabular*}{\textwidth}{c c}
\centering
&\includegraphics[width=3.2in]{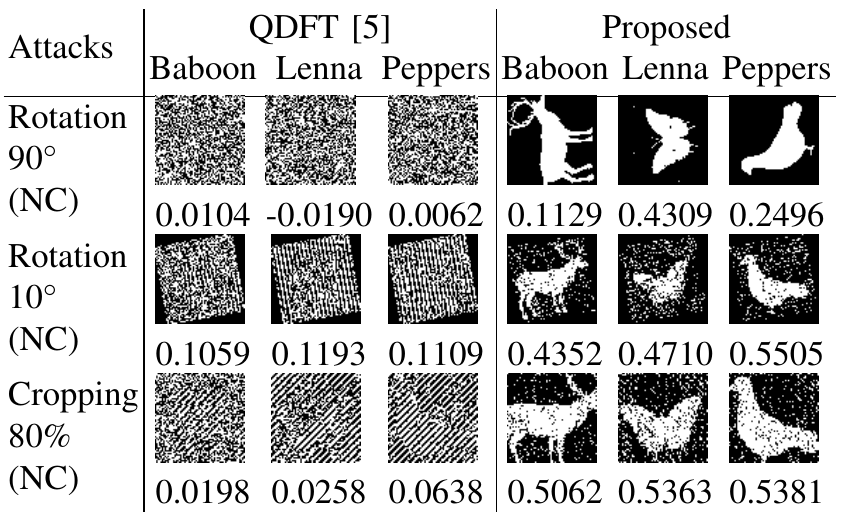}
\end{tabular*}
\end{table}

%\begin{figure}[]
%\centering
%\includegraphics[width=3.5in]{noreg}
%\caption{Changes in Cross-entropy loss and invisibility according to the embedding step. The points are determined by the average of the 128 blocks. A description of the PSNR is given in Section \ref{exp_results}.}
%\label{noreg}
%\end{figure}

The first CNN-based watermarking technique is proposed by \cite{kandi2017exploring}. Unlike the proposed technique, \cite{kandi2017exploring} uses CNN as an auto-encoder to present a new domain. However,  \cite{kandi2017exploring} is a non-blind method. Therefore, although the robustness of these techniques is high, they are less useful in real scenarios. The proposed technique, on the other hand, is a blind watermarking.

\section{Conclusion}

We proposed a robust blind watermarking technique using CNN. Experiments have yielded higher PSNR and NC values than comparative methods for test images. The learned models outperformed QDFT with assuming registration. Traditional watermarking researches first design the transform, and then attack and measure performance. In contrast, the proposed scheme optimized the watermarking domain through attack functions. Above all, the scheme does not require detailed algorithms of each attack or expert knowledge to counter them. Our technique was adaptive to the attacks and required only one day of learning time. 

%We are continuing our research to increase the embedding and detection speed while at the same time allowing the proposed method to have a template matching method.

% or


\ifCLASSOPTIONcaptionsoff
  \newpage
\fi

\bibliographystyle{IEEEtran}
\bibliography{bare_jrnl}

\end{document}